\documentclass[aps,twocolumn,pra,superscriptaddress,amsmath,amssymb]{revtex4-2} 
\usepackage{dcolumn}
\usepackage{amsmath}
\usepackage{mathrsfs}
\usepackage{txfonts}
\usepackage{bm}
\usepackage[T1]{fontenc}
\usepackage{xspace}
\usepackage{ulem}
\usepackage{comment}
\usepackage{braket}
\usepackage{lipsum}
\usepackage{bbold}
\usepackage{mathtools}
\usepackage{placeins}
\usepackage{graphicx}
\usepackage{hyperref}
\usepackage{color}
\usepackage{xcolor}
\usepackage[version=3]{mhchem}
\hypersetup{
        colorlinks=true,
        citecolor=blue,
        urlcolor=blue,
        linkcolor=blue
}

\begin{document}
\let\emph\textit

\title{
    Electronic Responses in Solid Oxygen with Coexisting Structural and Magnetic Orders 
}
\author{Naomichi Sato}
\affiliation{
  Department of Physics, Graduate School of Science, Tohoku University, Sendai, Miyagi 980-8578, Japan
}

\author{Makoto Naka}
\affiliation{
  School of Science and Engineering, Tokyo Denki University, Ishizaka 350-0394, Japan
}

\author{Joji Nasu}
\affiliation{
  Department of Physics, Graduate School of Science, Tohoku University, Sendai, Miyagi 980-8578, Japan
}

\date{\today}
\begin{abstract}
Solid oxygen provides a unique platform for studying the interplay between structural and magnetic degrees of freedom, as each O$_2$ molecule possesses an $S=1$ spin and 
weak intermolecular van der Waals interactions allow substantial structural flexibility.
In addition, because of its small intermolecular hopping integrals, this system is known to be a strongly correlated electron system with a multiorbital nature.
In this study, we examine electronic states and magnetic and dielectric responses using a multiorbital Hubbard model with Peierls-type electron-lattice coupling.
This model provides a minimal description of the interplay between structural and magnetic degrees of freedom in solid oxygen.
We apply the Hartree-Fock approximation to obtain the ground-state phase diagram.
Antiferromagnetic order appears when the onsite Coulomb interaction is strong,
whereas three nonmagnetic bond-ordered states emerge when it is weak. 
At intermediate interaction strengths, antiferromagnetic order coexists with each of the three bond orders.
These three coexistence phases all break time-reversal symmetry; one preserves spatial inversion symmetry, whereas the other two also break it.
The two inversion-breaking phases exhibit linear magnetoelectric responses, whereas the inversion-symmetric phase exhibits altermagnetic spin splitting and a piezomagnetic response.
These responses may provide experimental signatures of coexisting magnetic and structural orders in high-pressure solid oxygen suggested by recent experiments.
\end{abstract}
\maketitle

\section{Introduction}
\label{introduction}
In strongly correlated electron systems, the interplay among spin, charge, and orbital degrees of freedom is a key ingredient in stabilizing complex ordered phases and generating unconventional electronic responses.
Coupling these electronic degrees of freedom to the lattice can further drive structural distortions and qualitatively modify the physical responses.
This coupling is particularly pronounced in molecular solids, where relatively small intermolecular hopping integrals enhance correlation effects and weak intermolecular van der Waals interactions allow molecular electronic states to strongly influence the crystal structure.
Solid oxygen is a prototypical example: each $\mathrm O_2$ molecule possesses an $S=1$ spin in the doubly degenerate $\pi_g$ orbitals, and the molecular orientations and intermolecular arrangements are flexible.
These characteristics give rise to strong spin-lattice coupling and a pronounced dependence of both crystal structure and magnetic order on pressure and ultrahigh magnetic fields~\cite{Freiman2004,Freiman2018,Nomura2022}.

Thus far, pressure effects on solid oxygen have been extensively studied, and a variety of pressure-induced phases have been identified.
At ambient pressure and below approximately $24$ K, antiferromagnetic (AFM)
order appears in a monoclinic structure known as the $\alpha$
phase~\cite{Collins1966,Nomura2017}.
With increasing pressure, solid oxygen forms the orthorhombic $\delta$ phase at
pressures between approximately $5.5$ and $8$ GPa, which also exhibits
long-range AFM order~\cite{Gorelli2002,Goncharenko2004}.
Above approximately $8$ GPa, solid oxygen forms another monoclinic phase,
termed the $\varepsilon$ phase, in which long-range AFM order
disappears~\cite{Goncharenko2005}.
Detailed structural analyses have revealed that the $\varepsilon$ phase consists of $(\mathrm O_2)_4$ molecular units, in which four $\mathrm O_2$ molecules form a rhombus-like tetramer~\cite{Lundegaard2006,Fujihisa2006}.
This tetramerization can be regarded as a bond-ordered state characterized by a spatial pattern of inequivalent intermolecular bonds.
Furthermore, recent studies using ultrahigh magnetic fields above $100$ T
have reported a field-induced structural phase transition and giant
anisotropic magnetostriction in solid oxygen~\cite{Nomura2014,Ikeda2025}.
These observations provide experimental evidence that magnetic-field-induced changes in spin arrangements are closely linked to structural changes in solid oxygen.

Among the pressure-induced phases, the $\varepsilon$ phase has attracted particular attention because of the possible interplay between molecular magnetism and the bond order associated with the tetramerization.
While the magnetic and structural properties of the low-pressure $\alpha$
phase are well described by first-principles and many-body cluster
calculations~\cite{Jansen1987,daSilva1995,Obata2015,kasamatsu2017},
those of the $\varepsilon$ phase remain under debate.
Several spin-polarized first-principles calculations have found magnetic
solutions to be energetically favorable in the lower-pressure region of the
$\varepsilon$ phase, suggesting that molecular $S=1$ moments may survive in
the tetramerized structure~\cite{ANHPHAM2009160,Crespo2014,RamirezSolis2017,Anh2019}.
More specifically, Crespo \textit{et al.} proposed an $\varepsilon_1$ regime
at approximately $8$--$20$ GPa, in which the molecular moments form a
collective singlet state dominated by short-range AFM correlations, whereas
the higher-pressure $\varepsilon_0$ regime is nonmagnetic~\cite{Crespo2014}.
Subsequent first-principles calculations incorporating long-range van der
Waals interactions found spin-polarized states that were both energetically
stable and consistent with the experimentally observed structure~\cite{Anh2019}.
Recently, single-crystal x-ray diffraction measurements have revealed an isostructural transition at $18.1\pm0.5$ GPa associated with discontinuities in the lattice parameters and intratetramer distances~\cite{Gorelli2025}.
This observation has been interpreted as structural evidence for the spin collapse from the $\varepsilon_1$ state with molecular $S=1$ moments to the nonmagnetic $\varepsilon_0$ state.
These theoretical and experimental results support the possibility that
tetramerization coexists with molecular magnetism in the lower-pressure region
of the $\varepsilon$ phase.

However, it remains unclear what microscopic mechanisms underlie the
magnetic-to-nonmagnetic transition within the tetramerized state and how
tetramerization competes with other possible bond-order patterns.
Furthermore, the coexistence of magnetic order with different bond orders can
break time-reversal and/or spatial-inversion symmetry, giving rise to
cross-correlated responses involving spin, charge, and lattice degrees of
freedom.
Such responses can serve as experimental probes for identifying the underlying
magnetic and bond-order configurations.

In this study, we construct a minimal multiorbital model for solid oxygen in
which electrons are coupled to O$_2$ molecular displacements.
Using this model, we investigate the competition and coexistence between
magnetic and bond orders, as well as the associated electronic responses.
We employ a three-orbital Peierls--Hubbard model for two $\pi_g$ orbitals and a $\sigma_g$ orbital of an $\mathrm O_2$ molecule, where bond-dependent intermolecular hoppings and the intramolecular spin-orbit coupling are taken into account.
Applying the Hartree--Fock approximation, we obtain a ground-state phase
diagram containing an AFM phase and three bond-ordered phases with zigzag,
meander, and tetramer patterns.
At intermediate Coulomb interaction strengths, AFM order coexists with each
bond order, giving rise to three coexistence phases.
The coexistence phases with zigzag and meander bond orders break spatial
inversion symmetry ($\mathcal P$) and time-reversal symmetry ($\mathcal T$) while
preserving the combined $\mathcal{PT}$ symmetry, and exhibit linear
magnetoelectric responses.
In contrast, the tetramerized state with AFM order preserves $\mathcal P$ but
breaks $\mathcal{PT}$ symmetry, leading to nonrelativistic spin splitting characteristic
of an altermagnetic state and a piezomagnetic response.

This paper is organized as follows.
In the next section, we introduce the model and calculation method.
In Sec.~\ref{sec:Model_Hamiltonian}, we introduce the three-orbital Peierls--Hubbard model to address the magnetic and structural properties of solid oxygen.
In Sec.~\ref{sec:Hartree-Fock_method}, we explain the Hartree-Fock approximation to treat the Coulomb interactions and show how to characterize bond orders using the lattice-distortion variables.
In Sec.~\ref{sec:Results}, we present the results of the Hartree-Fock calculations.
Section~\ref{sec:PhaseDiagram} shows the phase diagram and classifies the symmetry of each coexistence phase.
In Sec.~\ref{sec:MagnetoelectricResponse}, we calculate the magnetoelectric responses of the phases in which AFM order coexists with zigzag or meander bond order.
In Sec.~\ref{sec:Altermagnet}, we demonstrate the altermagnetic character of the coexisting phase with AFM order and tetramerization through spin splitting in its electronic band structure.
In Sec.~\ref{sec:Discussion}, we discuss the relation between the obtained coexistence phases and the pressure-induced phases of solid oxygen.
Finally, Sec.~\ref{sec:Summary} is devoted to the summary.

\section{Model and Method}
\label{sec:Model}
In this section, we introduce the model Hamiltonian and 
the Hartree-Fock procedure used to classify the ordered phases.

\subsection{Model Hamiltonian}
\label{sec:Model_Hamiltonian}

\begin{figure}[tbp]
  \centering
      \includegraphics[width=1.0\columnwidth,clip]{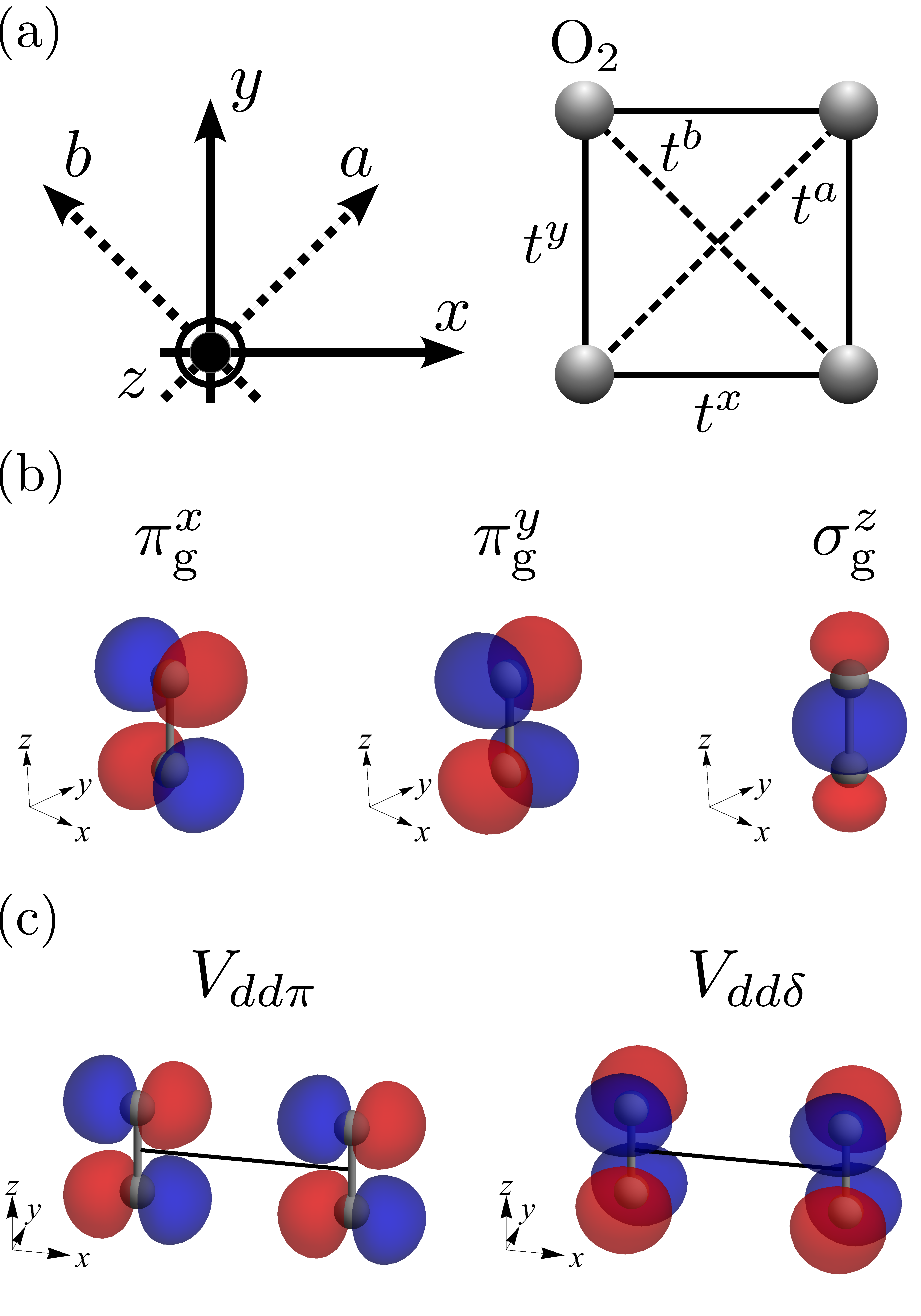}
      \caption{
        (a) Square-lattice representation of the molecular layer, in which each site
        represents an ${\rm O_2}$ molecule.
        The arrows define the $x$, $y$, $a$, $b$, and $z$ directions, where $z$ is
        perpendicular to the molecular layer.
        The hopping matrices $t^x$, $t^y$, $t^a$, and $t^b$ for the four in-plane
        bond directions are defined in Eq.~\eqref{eq:hopping_matrices}.
        The solid and dashed lines indicate nearest-neighbor and
        next-nearest-neighbor bonds, respectively.
        (b) Schematic pictures of the $\pi_g^x$, $\pi_g^y$, and $\sigma_g^z$
        molecular orbitals retained in the model.
        The molecular axis is taken along the $z$ direction, and the red and blue
        lobes represent the positive and negative signs of the molecular-orbital
        wave functions, respectively.
        (c) Relative configurations of the $\pi_g$ orbitals on neighboring ${\rm O_2}$ molecules corresponding to the $V_{dd\pi}$ and $V_{dd\delta}$ hopping integrals.
      }
      \label{fig:model_transfer}
\end{figure}

Since the tetramerization characteristic of the $\varepsilon$ phase mainly develops within the crystallographic $ab$ plane, we focus on an O$_2$ molecular layer in this plane and construct a two-dimensional model. Although the $\alpha$ and $\varepsilon$ phases have monoclinic crystal structures, for simplicity, we neglect the in-plane crystallographic anisotropy and approximate the molecular arrangement by a square lattice.
We take the $x$ and $y$ axes along the nearest-neighbor bond directions and
the crystallographic $a$ and $b$ axes along the next-nearest-neighbor bond
directions.
The $z$ axis is taken along the molecular axis perpendicular to the $ab$
plane, as shown in Fig.~\ref{fig:model_transfer}(a).
On this lattice, we consider a three-orbital Peierls--Hubbard model that
includes the degenerate highest occupied molecular orbitals $\pi_g^x$ and
$\pi_g^y$ and the occupied $\sigma_g^z$ orbital.
The latter orbital is introduced to describe the magnetic anisotropy induced by
spin-orbit coupling, as discussed below.
The model includes lattice-distortion degrees of freedom that modulate the
hopping amplitudes and describe bond alternation.
The Hamiltonian is given by
\begin{equation}
  \label{eq:Hamiltonian}
  \mathcal H
  =
  \mathcal H_{U}
  +
  \mathcal H_{\rm orb}
  +
  \mathcal H_{\rm SOC}
  +
  \mathcal H_{t}
  +
  \mathcal H_{K},
\end{equation}
where $\mathcal H_{U}$ describes the onsite Coulomb interactions,
$\mathcal H_{\rm orb}$ represents the molecular orbital energy levels,
$\mathcal H_{\rm SOC}$ represents the spin-orbit coupling,
$\mathcal H_{t}$ accounts for electron hopping between neighboring and next-nearest-neighbor molecules, and
$\mathcal H_{K}$ describes the elastic energy associated with lattice distortions.

The onsite Coulomb interaction term $\mathcal H_{U}$ is expressed as
\begin{align}
  \label{eq:H_U}
  \mathcal H_{U}
  =
  U\sum_{i}\sum_{\mu}n_{i\mu\uparrow}n_{i\mu\downarrow}
  +
  U'\sum_{i}\sum_{\mu>\nu}n_{i\mu}n_{i\nu} \nonumber \\
  +
  J_{\rm H}
  \sum_{i}\sum_{\mu>\nu}\sum_{ss'}
  c^{\dagger}_{i\mu s}c^{\dagger}_{i\nu s'}
  c_{i\mu s'}c_{i\nu s},
\end{align}
where $c^{\dagger}_{i\mu s}$ and $c_{i\mu s}$ are the creation and
annihilation operators of an electron with spin $s$ in orbital $\mu$ on
the $i$-th ${\rm O_2}$ molecule, respectively.
The corresponding number operator is defined as
$n_{i\mu s}=c^{\dagger}_{i\mu s}c_{i\mu s}$, and
$n_{i\mu}=\sum_{s}n_{i\mu s}$.
The orbital indices $\mu$ and $\nu$ take the values
$\pi_g^x$, $\pi_g^y$, and $\sigma_g^z$.
Their spatial forms are schematically illustrated in
Fig.~\ref{fig:model_transfer}(b).
Here, $U$ and $U'$ are the intra- and inter-orbital Coulomb repulsions,
respectively, and $J_{\rm H}$ is the Hund's coupling.
Pair-hopping terms are neglected in the present model.
The ratios among these interaction parameters are fixed as
$U:U':J_{\rm H} = 11.6:11.4:0.80$, based on optical absorption measurements 
of solid oxygen~\cite{LANDAU19621} and first-principles calculations~\cite{kasamatsu2017}.

The molecular-orbital energy term $\mathcal H_{\rm orb}$ is given by
\begin{equation}
  \label{eq:H_orb}
  \mathcal H_{\rm orb}
  =
  -E
  \sum_{i,s}
  n_{i,\sigma_g^z,s},
\end{equation}
where $E$ denotes the energy offset of the $\sigma_g^z$ molecular orbital
relative to the $\pi_g$ orbitals.
The spin-orbit coupling term $\mathcal H_{\rm SOC}$ is given by
\begin{equation}
  \label{eq:H_SOC}
  \mathcal H_{\rm SOC}
  =
  \lambda
  \sum_i
  \sum_{\mu\nu}
  \sum_{ss'}
  c^\dagger_{i\mu s}
  \left(
  \bm{\ell}\cdot\bm{s}
  \right)_{\mu s,\nu s'}
  c_{i\nu s'},
\end{equation}
where $\lambda$ is the spin-orbit coupling constant.
The matrices $\bm{\ell}$ and $\bm{s}$ represent the orbital and spin angular
momentum operators, respectively, in units of $\hbar$. 
The spin matrices are given by $\bm{s}=\bm{\sigma}/2$, 
where $\bm{\sigma}$ is the vector of Pauli matrices.
In the orbital basis
$\{|\pi_g^x\rangle,|\pi_g^y\rangle,|\sigma_g^z\rangle\}$,
the components of $\bm{\ell}$ are represented as
\begin{equation}
  \label{eq:orbital_angular_momentum}
  \ell^x
  =
  \begin{pmatrix}
    0 & 0 & 0 \\
    0 & 0 & -i \\
    0 & i & 0
  \end{pmatrix},
  \quad
  \ell^y
  =
  \begin{pmatrix}
    0 & 0 & i \\
    0 & 0 & 0 \\
    -i & 0 & 0
  \end{pmatrix},
  \quad
  \ell^z
  =
  \begin{pmatrix}
    0 & -i & 0 \\
    i & 0 & 0 \\
    0 & 0 & 0
  \end{pmatrix}.
\end{equation}
Here, we evaluate the orbital angular momentum matrix elements of ${\rm O_2}$
within the one-center approximation.
This approximation retains only the onsite atomic $2p$ contributions and
neglects the interatomic ones, following previous treatments of molecular
oxygen and related species~\cite{Minaev1996,Prabhakar2002}.
Details of this approximation, including a derivation of
Eq.~\eqref{eq:orbital_angular_momentum}, are provided in
Appendix~\ref{app:orbital_angular_momentum}.
This term accounts for the in-plane magnetic anisotropy of solid oxygen
~\cite{Obata2018}.

The hopping term $\mathcal H_{t}$ is expressed as
\begin{equation}
  \label{eq:H_t}
  \mathcal H_t
  =
  -
  \sum_{(i,j)\in \mathcal B}
  \sum_{\mu,\nu}
  \sum_{s}
  \left(
  1+\delta_{ij}
  \right)
  t^{ij}_{\mu\nu}
  c^{\dagger}_{i\mu s}
  c_{j\nu s}
  +{\rm H.c.}.
\end{equation}
Here, $\mathcal B$ denotes the set of molecular bonds included in the hopping
term, consisting of the nearest-neighbor bond set $\mathcal B_{\rm NN}$ and the
next-nearest-neighbor bond set $\mathcal B_{\rm NNN}$.
For both nearest-neighbor and next-nearest-neighbor bonds, the signed
dimensionless Peierls modulation $\delta_{ij}$ is defined by
\begin{equation}
  \label{eq:peierls_modulation}
  \delta_{ij}
  =
  \alpha
  \hat{\bm e}_{ij}\cdot
  \left(
  \bm u_i-\bm u_j
  \right),
\end{equation}
where $\alpha$ is the electron-lattice coupling constant,
$\hat{\bm e}_{ij}$ is the unit vector along the bond from molecule $j$ to
molecule $i$, and $\bm u_i$ is the displacement of molecule $i$.
The hopping matrix $t^{ij}_{\mu\nu}$ depends on the bond direction.
In the hopping term, we retain only the matrix elements within the
$\pi_g^x$ and $\pi_g^y$ orbitals, neglecting the $\sigma_g^z$ orbital.
As illustrated in Fig.~\ref{fig:model_transfer}(c), $V_{dd\pi}$ and
$V_{dd\delta}$ denote the nearest-neighbor Slater--Koster hopping parameters
for the $dd\pi$ and $dd\delta$ processes, respectively
~\cite{SlaterKoster1954}.
The corresponding next-nearest-neighbor hopping integrals are denoted by
$W_{dd\pi}$ and $W_{dd\delta}$.
For notational simplicity, we introduce
$W_{\pm}=(W_{dd\pi}\pm W_{dd\delta})/2$.
In the orbital basis
$\{|\pi_g^x\rangle,|\pi_g^y\rangle\}$,
the hopping matrices for the four bond directions are defined as
\begin{equation}
  \label{eq:hopping_matrices}
  \begin{alignedat}{2}
  t^x
  &=
  \begin{pmatrix}
    V_{dd\pi} & 0 \\
    0 & V_{dd\delta}
  \end{pmatrix},
  &\qquad
  t^y
  &=
  \begin{pmatrix}
    V_{dd\delta} & 0 \\
    0 & V_{dd\pi}
  \end{pmatrix},
  \\
  t^a
  &=
  \begin{pmatrix}
    W_{+} & W_{-} \\
    W_{-} & W_{+}
  \end{pmatrix},
  &\qquad
  t^b
  &=
  \begin{pmatrix}
    W_{+} & -W_{-} \\
    -W_{-} & W_{+}
  \end{pmatrix}.
  \end{alignedat}
\end{equation}
The matrices $t^x$ and $t^y$ correspond to the nearest-neighbor bonds along
the $x$ and $y$ directions, respectively, whereas $t^a$ and $t^b$ correspond
to the next-nearest-neighbor bonds along the $a$ and $b$ directions.
The corresponding bond directions are illustrated in Fig.~\ref{fig:model_transfer}(a).
We assume that the next-nearest-neighbor hopping integrals are obtained by
scaling their nearest-neighbor counterparts by the same factor $w$, so that
\begin{equation}
  \label{eq:w_definition}
  w
  =
  \frac{W_{dd\pi}}{V_{dd\pi}}
  =
  \frac{W_{dd\delta}}{V_{dd\delta}},
\end{equation}
and consequently
$W_{\pm}=w(V_{dd\pi}\pm V_{dd\delta})/2$.
We take $V_{dd\pi}>0$ as the unit of energy, set $V_{dd\pi}=1$, and fix
$V_{dd\delta}/V_{dd\pi}=-0.01$ in the numerical calculations.
For each bond $(i,j)$, $t^{ij}_{\mu\nu}$ in Eq.~\eqref{eq:H_t} is given by
the $(\mu,\nu)$ element of the corresponding matrix $t^x$, $t^y$, $t^a$ or
$t^b$.

\begin{figure}[t]
  \centering
      \includegraphics[width=1.0\columnwidth,clip]{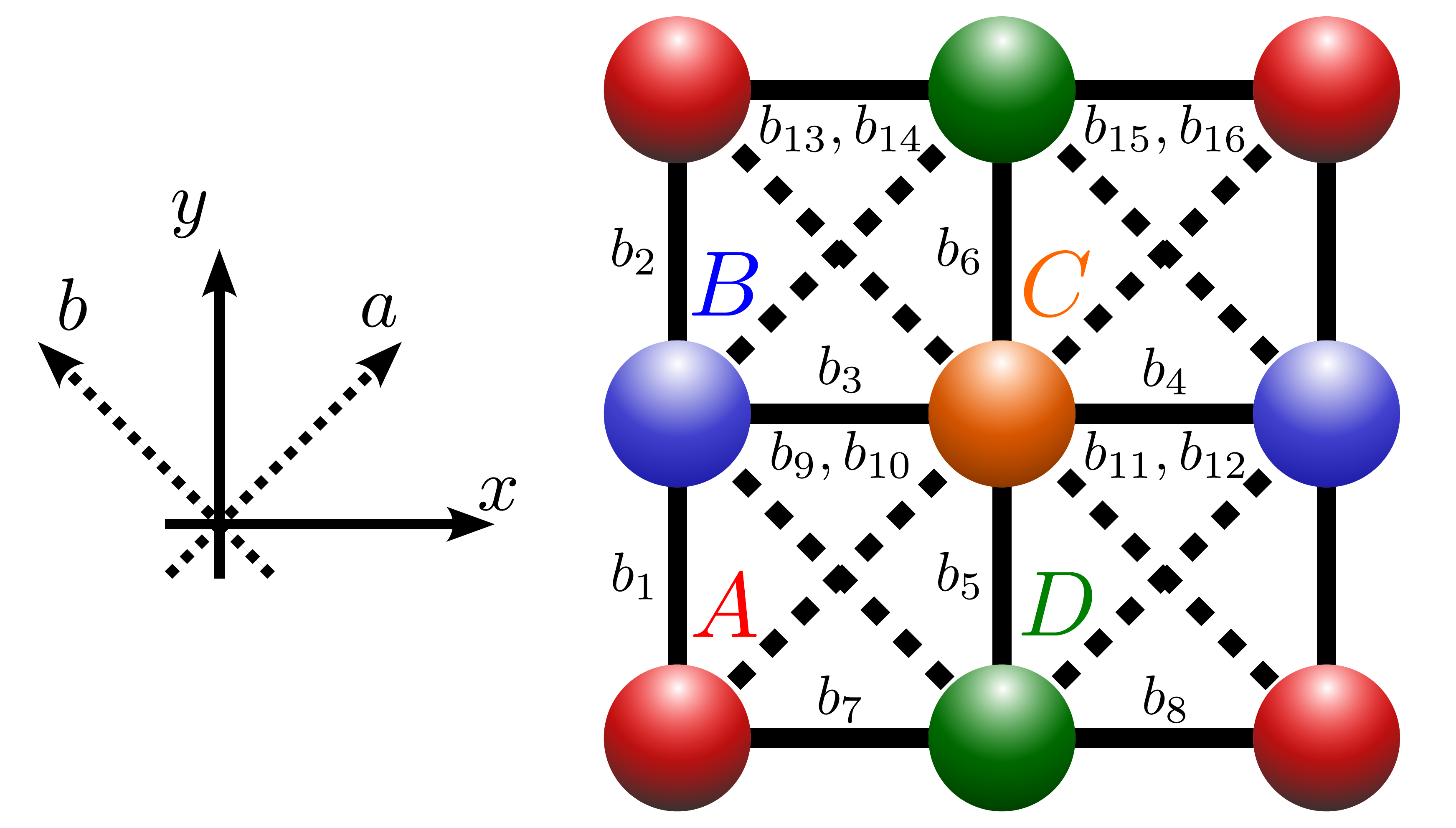}
      \caption{
        Definition of the four sublattices and independent molecular bonds in the four-sublattice unit cell.
        The arrows on the left define the $x$, $y$, $a$, and $b$ directions.
        The nearest-neighbor and next-nearest-neighbor bonds are indicated by solid and dashed lines, respectively. 
        }
      \label{fig:model_bond}
\end{figure}

To describe the bond alternation associated with the structural transition,
we assume a four-sublattice unit cell consisting of sublattices
$A$, $B$, $C$, and $D$, as shown in Fig.~\ref{fig:model_bond}.
This choice allows us to include the lattice distortions associated with
the tetramerization observed in the $\varepsilon$ phase of solid oxygen.
We choose four independent parameters to describe the distortions of the
nearest-neighbor bonds as
\begin{align}
  \label{eq:delta_NN}
  \delta_{AB}
  &=
  \alpha
  \hat{\bm e}_{y}\cdot
  \left(
  \bm u_A-\bm u_B
  \right),
  \quad
  \delta_{BC}
  =
  \alpha
  \hat{\bm e}_{x}\cdot
  \left(
  \bm u_B-\bm u_C
  \right),
  \nonumber\\
  \delta_{DC}
  &=
  \alpha
  \hat{\bm e}_{y}\cdot
  \left(
  \bm u_D-\bm u_C
  \right),
  \quad
  \delta_{AD}
  =
  \alpha
  \hat{\bm e}_{x}\cdot
  \left(
  \bm u_A-\bm u_D
  \right),
\end{align}
which correspond to the $AB$, $BC$, $DC$, and $AD$ bonds, respectively.
For each nearest-neighbor bond, the signed modulation $\delta_{ij}$ is
identified with one of $\pm\delta_{AB}$, $\pm\delta_{BC}$,
$\pm\delta_{DC}$, and $\pm\delta_{AD}$ according to its bond class.
Because the next-nearest-neighbor bond distortions are determined by the same
four sublattice displacements, their Peierls modulation factors can be
expressed in terms of the four nearest-neighbor variables in
Eq.~\eqref{eq:delta_NN}. 
For the next-nearest-neighbor bonds, we introduce two modulation variables, 
$\zeta_1$ and $\zeta_2$, defined by
\begin{align}
  \label{eq:delta_NNN}
  \zeta_{1}
  &= \frac{1}{2\sqrt{2}}\left(\delta_{AB} + \delta_{BC} + \delta_{DC} + \delta_{AD}\right), \\
  \zeta_{2}
  &= \frac{1}{2\sqrt{2}}\left(\delta_{AB} - \delta_{BC} + \delta_{DC} - \delta_{AD}\right).
\end{align}
These variables represent the distortion components entering the
next-nearest-neighbor Peierls modulations.
The signed modulation $\delta_{ij}$ for each next-nearest-neighbor bond is
identified with one of $\pm\zeta_1$ and $\pm\zeta_2$.
Thus, the inclusion of next-nearest-neighbor hopping does not introduce
additional independent lattice-distortion variables in the present model.
The hopping matrices and Peierls modulations for all nearest-neighbor and
next-nearest-neighbor bonds are summarized in
Table~\ref{tab:bond_definition}.

\begin{table}[tbp]
\caption{Hopping matrices and bond-dependent Peierls modulation variables for
the nearest-neighbor and next-nearest-neighbor molecular bonds.
The bond labels correspond to those shown in Fig.~\ref{fig:model_bond}.
For each bond, the hopping matrix is multiplied by
$1+\delta_{ij}$.}
\begin{ruledtabular}
\begin{tabular}{ccccc}
Bond & $\mathcal B$ & $(i,j)$ & Hopping matrix & $\delta_{ij}$ \\
\hline
$b_1$ & $\mathcal B_{\rm NN}$ & $(A,B)$ & $t^y$ & $+\delta_{AB}$ \\
$b_2$ & $\mathcal B_{\rm NN}$ & $(A,B)$ & $t^y$ & $-\delta_{AB}$ \\
$b_3$ & $\mathcal B_{\rm NN}$ & $(B,C)$ & $t^x$ & $+\delta_{BC}$ \\
$b_4$ & $\mathcal B_{\rm NN}$ & $(B,C)$ & $t^x$ & $-\delta_{BC}$ \\
$b_5$ & $\mathcal B_{\rm NN}$ & $(C,D)$ & $t^y$ & $+\delta_{DC}$ \\
$b_6$ & $\mathcal B_{\rm NN}$ & $(C,D)$ & $t^y$ & $-\delta_{DC}$ \\
$b_7$ & $\mathcal B_{\rm NN}$ & $(D,A)$ & $t^x$ & $+\delta_{AD}$ \\
$b_8$ & $\mathcal B_{\rm NN}$ & $(D,A)$ & $t^x$ & $-\delta_{AD}$ \\
$b_9$  & $\mathcal B_{\rm NNN}$ & $(B,D)$ & $t^b$ & $+\zeta_{1}$ \\
$b_{10}$ & $\mathcal B_{\rm NNN}$ & $(A,C)$ & $t^a$ & $+\zeta_{1}$ \\
$b_{11}$ & $\mathcal B_{\rm NNN}$ & $(A,C)$ & $t^b$ & $+\zeta_{2}$ \\
$b_{12}$ & $\mathcal B_{\rm NNN}$ & $(B,D)$ & $t^a$ & $+\zeta_{2}$ \\
$b_{13}$ & $\mathcal B_{\rm NNN}$ & $(A,C)$ & $t^b$ & $-\zeta_{2}$ \\
$b_{14}$ & $\mathcal B_{\rm NNN}$ & $(B,D)$ & $t^a$ & $-\zeta_{2}$ \\
$b_{15}$ & $\mathcal B_{\rm NNN}$ & $(B,D)$ & $t^b$ & $-\zeta_{1}$ \\
$b_{16}$ & $\mathcal B_{\rm NNN}$ & $(A,C)$ & $t^a$ & $-\zeta_{1}$ \\
\end{tabular}
\end{ruledtabular}
\label{tab:bond_definition}
\end{table}

Using Eq.~\eqref{eq:delta_NN} and noting that each of the four
nearest-neighbor bond distortions contributes twice to the elastic energy
per unit cell, we obtain
\begin{equation}
  \label{eq:H_K_lambda}
  \mathcal H_K
  =
  N_{\mathrm{cell}}
  \frac{V_{dd\pi}}{\eta}
  \left(
  \delta_{AB}^{2}
  +
  \delta_{BC}^{2}
  +
  \delta_{DC}^{2}
  +
  \delta_{AD}^{2}
  \right),
\end{equation}
where $\eta=\alpha^2V_{dd\pi}/K$ is the dimensionless electron-lattice
coupling strength, $K$ is the elastic constant, and $N_{\mathrm{cell}}$
is the number of four-sublattice unit cells.
We fix $\eta=0.90$ throughout the calculations.

\subsection{Hartree-Fock method and order parameters}
\label{sec:Hartree-Fock_method}
To solve the Hamiltonian in Eq.~\eqref{eq:Hamiltonian}, we employ the
Hartree-Fock approximation, in which the interaction terms in
Eq.~\eqref{eq:H_U} are decoupled into one-body terms.
The mean fields are represented by the local density matrix
$\langle c^\dagger_{i\mu s}c_{i\nu s'}\rangle$ and allowed to
depend on the four sublattices shown in Fig.~\ref{fig:model_bond}.
In performing this decoupling, we include all one-body mean fields that
conserve the total particle number and determine them self-consistently.
The calculations are performed at zero temperature for four electrons per site,
corresponding to two electrons distributed over the $\pi_g^x$ and $\pi_g^y$
orbitals and two electrons occupying the $\sigma_g^z$ orbital of each
${\rm O_2}$ molecule.
The nearest-neighbor intermolecular distance is taken as the unit of length.
The lattice distortions in Eq.~\eqref{eq:delta_NN} are determined
self-consistently from the stationary conditions for the total energy, which
are evaluated using the Hellmann-Feynman theorem.


After obtaining the self-consistent Hartree-Fock solutions, we evaluate the
magnetic and bond order parameters defined below to classify the resulting
phases.
The magnetic order parameter is defined as the Fourier component of the
sublattice spin moments along the $b$ axis,
\begin{align}
  \label{eq:spin_structure_factor_momentum}
  S_b(\bm q)
  &=
  \frac{1}{4}
  \sum_{\gamma=A,B,C,D}
  m_{\gamma}^{b}
  e^{i\bm q\cdot\bm R_{\gamma}},\\
  \label{eq:spin_structure_factor_real}
  m_{\gamma}^{b}
  &=
  \frac{1}{2}
  \sum_{\mu}
  \sum_{ss'}
  \langle c^{\dagger}_{\gamma\mu s}
  \sigma^{b}_{ss'}
  c_{\gamma\mu s'}\rangle,
\end{align}
where $m_{\gamma}^{b}$ is the $b$-axis component of the spin moment on
sublattice $\gamma$ and $\bm R_\gamma$ is its position within the unit cell.
The matrix $\sigma^b=-(\sigma^x-\sigma^y)/\sqrt{2}$ is the Pauli matrix along
the $b$ axis.
In the following analysis, we mainly use the components at
$\bm q=(0,0)$ and $(\pi,\pi)$, which characterize the ferromagnetic and
antiferromagnetic components within the four-sublattice unit cell,
respectively.
We also characterize the bond-order patterns using the four nearest-neighbor
bond distortions in Eq.~\eqref{eq:delta_NN}. 
The bond-order parameters are constructed from products of distortions on
opposite nearest-neighbor bonds.  

\begin{figure}[t]
  \centering
      \includegraphics[width=1.0\columnwidth,clip]{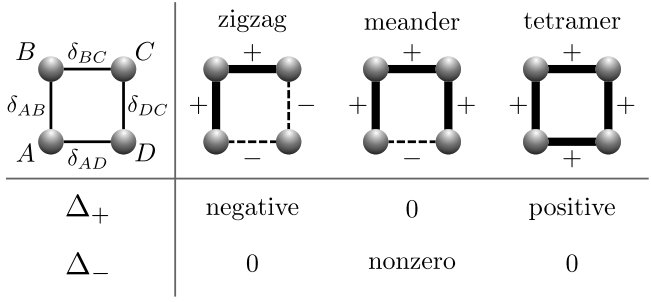}
      \caption{
        Schematic relation between the nearest-neighbor bond distortions defined in
        Eq.~\eqref{eq:delta_NN} and the combinations $\Delta_{\pm}$ defined in
        Eq.~\eqref{eq:Delta_pm}.
        The diagram on the left shows the four independent bond distortions
        $\delta_{AB}$, $\delta_{BC}$, $\delta_{DC}$, and $\delta_{AD}$.
        The three diagrams on the right show the zigzag, meander, and tetramer
        distortion patterns.
        Solid and dashed bonds, together with the plus and minus signs, indicate
        positive and negative bond distortions, respectively.
        The values of $\Delta_{+}$ and $\Delta_{-}$ corresponding to each pattern
        are indicated below the diagrams.
      }
      \label{fig:model_orderparameter}
\end{figure}

We introduce the symmetric and antisymmetric
combinations
\begin{align}
  \label{eq:Delta_pm}
  \Delta_{+}
  &=
  \frac{1}{2}
  \left(
  \delta_{AB}\delta_{DC}
  +
  \delta_{BC}\delta_{AD}
  \right),
  &
  \Delta_{-}
  &=
  \frac{1}{2}
  \left(
  \delta_{AB}\delta_{DC}
  -
  \delta_{BC}\delta_{AD}
  \right),
\end{align}
and define the zigzag, meander, and tetramer bond-order parameters as
\begin{align}
  \label{eq:bond_order_parameters}
  \Delta_{\rm Z}
  &=
  {\rm Im}
  \left[
  \sqrt{\Delta_{+}}
  \right],
  &
  \Delta_{\rm M}
  &=
  \sqrt{
  \left|
  \Delta_{-}
  \right|
  },
  &
  \Delta_{\rm T}
  &=
  {\rm Re}
  \left[
  \sqrt{\Delta_{+}}
  \right].
\end{align}
The correspondence between the four bond distortions and the three
bond-distortion patterns is illustrated schematically in
Fig.~\ref{fig:model_orderparameter}.
For the zigzag and tetramer patterns, $\Delta_{-}=0$, and the two patterns are
distinguished by the sign of $\Delta_{+}$: $\Delta_{+}<0$ for zigzag and
$\Delta_{+}>0$ for tetramer.
For the meander pattern, $\Delta_{+}=0$, whereas $\Delta_{-}$ is nonzero.
Consequently, only one of $\Delta_{\rm Z}$, $\Delta_{\rm M}$, and $\Delta_{\rm T}$
becomes nonzero for each of the zigzag, meander, and tetramer patterns.
These quantities can be used as mutually exclusive order parameters
to distinguish the three bond-ordered phases.

\section{Results}
\label{sec:Results}

\subsection{Phase diagram and symmetry classification}
\label{sec:PhaseDiagram}

We first present the Hartree-Fock ground-state phase diagram by varying
the relative next-nearest-neighbor hopping strength $w$ and the normalized
onsite Coulomb interaction $U/V_{dd\pi}$. We fix $E/V_{dd\pi}=5.0$ and set
the spin-orbit coupling to $\lambda=0$, since it mainly affects the spin
orientation and does not qualitatively alter the phase competition between 
magnetic and bond orders discussed here.

\begin{figure*}[tbp]
  \centering
      \includegraphics[width=1.8\columnwidth,clip]{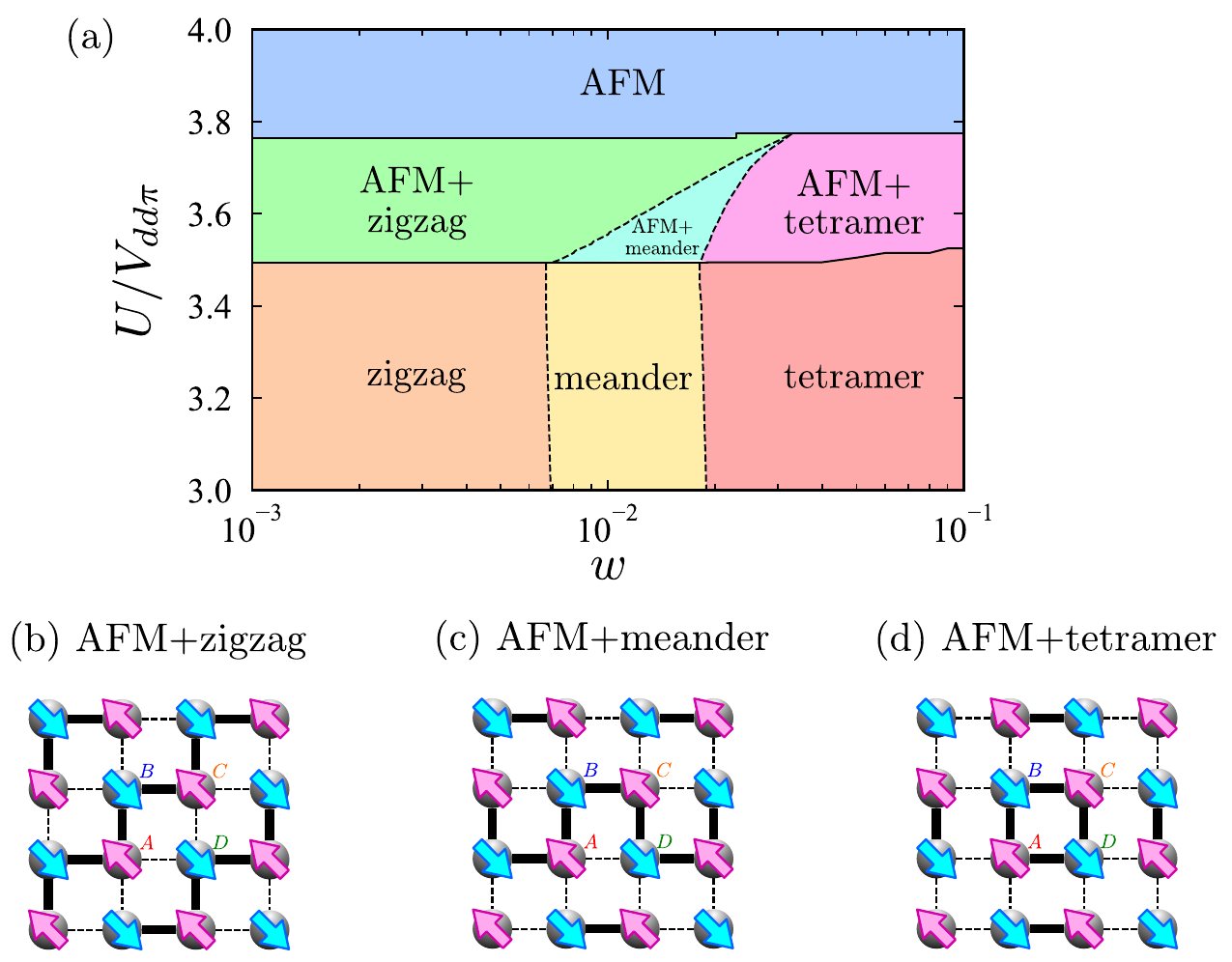}
      \caption{
        Ground-state phase diagram and representative real-space configurations of
        the coexistence phases.
        (a) Hartree-Fock phase diagram in the $w$-$U/V_{dd\pi}$ plane without
        spin-orbit coupling.
        Dashed and solid lines indicate first- and second-order phase boundaries,
        respectively.
        The spin and bond configurations are shown for
        (b) AFM+zigzag, (c) AFM+meander, and (d) AFM+tetramer phases.
        The arrows indicate the AFM spin configuration, and the solid and dashed
        bonds represent different nearest-neighbor bond distortions.
      }
      \label{fig:coexistence_phases}
\end{figure*}

Figure~\ref{fig:coexistence_phases}(a) shows the calculated ground-state phase
diagram in the $w$-$U$ plane.
In the large-$U$ region, the AFM phase is stabilized.
In the small-$U$ region, three bond-ordered phases appear depending on the
next-nearest-neighbor hopping strength $w$.  The zigzag, meander, and tetramer
bond orders are stabilized at small, intermediate, and large $w$,
respectively.
These bond orders are distinguished by the spatial patterns of the
nearest-neighbor bond distortions in the four-sublattice unit cell.
At intermediate $U$, AFM order coexists with each of the three bond orders.
Figures~\ref{fig:coexistence_phases}(b), \ref{fig:coexistence_phases}(c), and
\ref{fig:coexistence_phases}(d) show the schematic spin and bond configurations
of the three coexistence phases, which we refer to as AFM+zigzag, AFM+meander,
and AFM+tetramer, respectively.
The three coexistence phases share the same AFM spin pattern 
but differ in their bond-distortion patterns.
Since the spin-orbit coupling is not included in the phase-diagram calculation,
the spin direction is degenerate. 
We choose the spin direction parallel to the $b$ axis, 
following the experimentally observed spin orientation, and
use $S_b(\pi,\pi)$ to identify the AFM order.

\begin{figure*}[tbp]
  \centering
      \includegraphics[width=1.0\textwidth,clip]{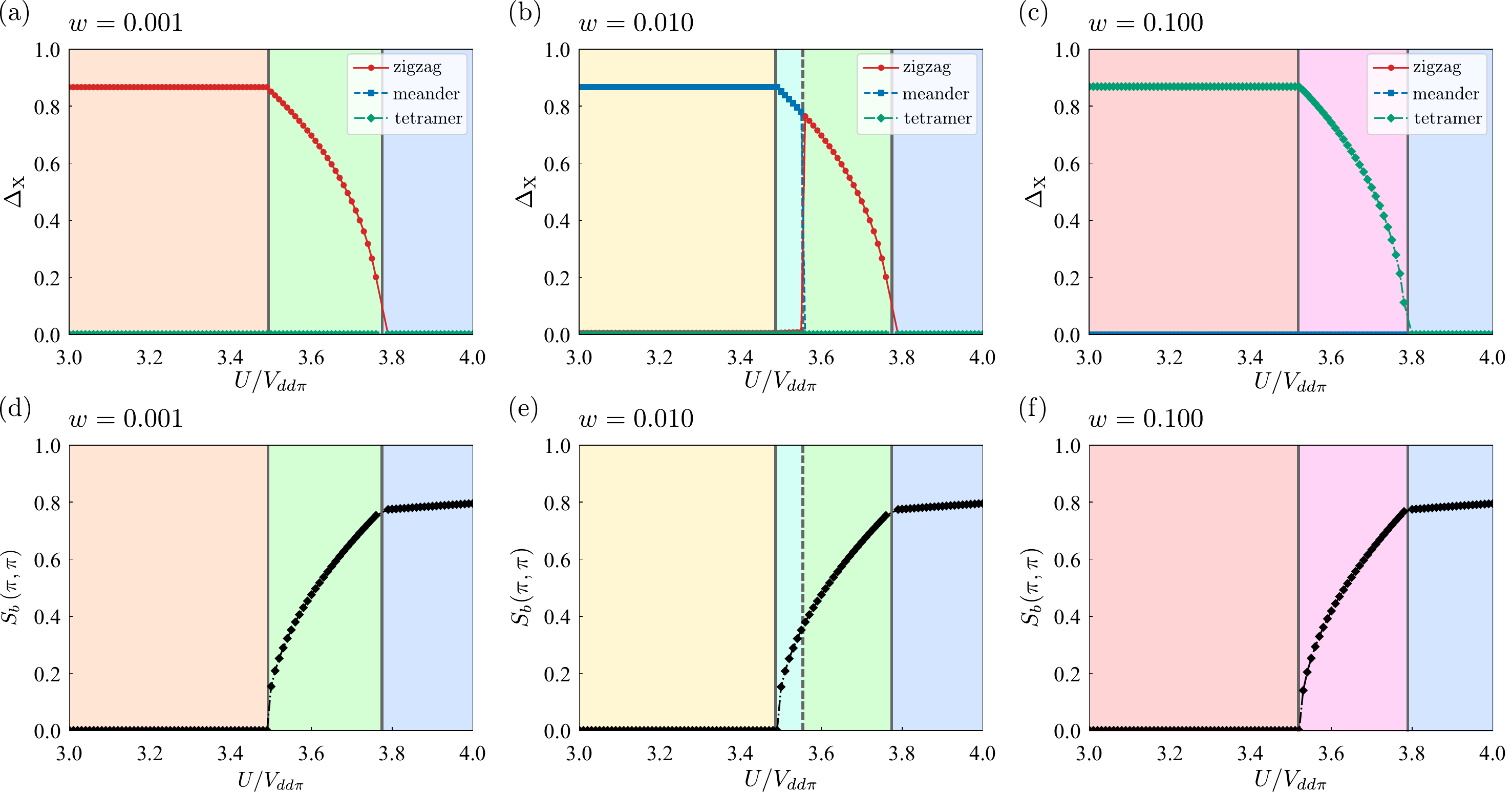}
      \caption{
        Bond and magnetic order parameters as functions of the onsite interaction
        $U/V_{dd\pi}$ for different next-nearest-neighbor hopping strengths
        $w=0.001$, $0.010$, and $0.100$.
        The upper row, (a)--(c), shows $\Delta_{\rm X}$ $(X={\rm Z},{\rm M},{\rm T})$, and
        the lower row, (d)--(f), shows $S_b(\pi,\pi)$.
        The pairs (a,d), (b,e), and (c,f) correspond to $w=0.001$, $0.010$, and
        $0.100$, respectively.
        The vertical dashed and solid lines indicate first- and second-order phase
        boundaries, respectively.
      }
      \label{fig:order_parameters}
\end{figure*}

To clarify how these phases are identified, we show representative cuts of
the phase diagram at $w=0.001$, $0.010$, and $0.100$ in
Fig.~\ref{fig:order_parameters}.
For all three values of $w$, the bond order parameter is finite in the
small-$U$ region, whereas the magnetic order parameter $S_b(\pi,\pi)$ vanishes.
With increasing $U/V_{dd\pi}$, $S_b(\pi,\pi)$ becomes finite, indicating the
onset of AFM order coexisting with bond order.  In the large-$U$ region, the
bond order disappears and only the AFM order remains.
The onset of AFM order and the disappearance of each bond order occur
continuously and are therefore regarded as second-order transitions
within the Hartree-Fock calculation.
In contrast, the transitions between different bond-distortion patterns are
accompanied by discontinuous changes in the bond order parameters and are
classified as first-order transitions.
For $w=0.001$, $0.010$, and $0.100$, the dominant bond orders are zigzag,
meander, and tetramer, respectively.
Previous studies using Peierls--Hubbard and related models have also reported
the coexistence of AFM order and bond distortion
~\cite{Yonemitsu1993,YuanKopp2002,Ly2025}.
Moreover, various bond-distortion patterns can be stabilized depending on
electronic interactions, the hopping structure, and elastic energy
~\cite{Tang1988,YuanNunnerKopp2001,Aits2006,Seo2002}.

We next classify the coexistence phases in terms of spatial inversion
symmetry $\mathcal P$, time-reversal symmetry $\mathcal T$, and their combined
symmetry $\mathcal{PT}$, which are relevant to the electronic responses
discussed below.
The symmetry properties of the AFM+zigzag, AFM+meander, and AFM+tetramer
phases are summarized in Table~\ref{tab:symmetry_coexistence}.

\begin{table}[tbp]
  \centering
  \caption{
    Symmetry properties of the coexistence phases.
  }
  \label{tab:symmetry_coexistence}
  \begin{tabular}{lccc}
    \hline\hline
    & AFM+zigzag & AFM+meander & AFM+tetramer \\
    \hline
    $\mathcal P$ symmetry   & $-$ & $-$ & $\checkmark$ \\
    $\mathcal T$ symmetry   & $-$ & $-$ & $-$ \\
    $\mathcal{PT}$ symmetry & $\checkmark$ & $\checkmark$ & $-$ \\
    \hline\hline
  \end{tabular}
\end{table}

The inversion symmetry depends on the type of bond order.
For the zigzag and meander distortions, the inversion center of the 
corresponding bond-order patterns is located at a bond center, 
while that of the AFM spin pattern is located at a
lattice site.  Their coexistence therefore breaks spatial inversion symmetry
$\mathcal P$.  In these phases, $\mathcal T$ symmetry is also broken.
Although $\mathcal P$ and $\mathcal T$ are individually broken, 
their combination $\mathcal{PT}$ remains a symmetry operation for the AFM+zigzag and
AFM+meander phases.
Since a linear magnetoelectric response requires both spatial inversion and time-reversal symmetries to be broken, these phases satisfy the necessary symmetry condition for the linear magnetoelectric effect~\cite{Dzyaloshinskii1960, Astrov1960}.

In contrast, for the tetramer distortion, the bond-order pattern and the AFM spin
pattern share an inversion center at the plaquette center.  Thus, the
AFM+tetramer phase preserves $\mathcal P$ symmetry.  However, $\mathcal T$
symmetry is broken, and consequently the combined $\mathcal{PT}$ symmetry is
also broken. 
This broken $\mathcal{PT}$ symmetry permits the spin splitting characteristic
of an altermagnetic phase, as examined below~\cite{Naka2019,Naka2025}.

Based on this symmetry classification, we calculate the magnetoelectric
response in the inversion-symmetry-broken AFM+zigzag and AFM+meander phases.
For the $\mathcal{PT}$-broken AFM+tetramer phase, we examine both the spin
splitting of the electronic band structure and the piezomagnetic response.
For these response calculations, we include finite spin-orbit coupling.
Although spin-orbit coupling does not qualitatively alter the phase
competition discussed above, it lifts the spin-rotation degeneracy and
fixes the magnetic easy axis, 
thereby allowing a symmetry-based analysis of these responses.

\subsection{Magnetoelectric response in AFM+zigzag and AFM+meander phases}
\label{sec:MagnetoelectricResponse}

We first examine the magnetoelectric responses of the AFM+zigzag and AFM+meander phases.
These phases break both spatial inversion symmetry $\mathcal P$ and time-reversal
symmetry $\mathcal T$.
To apply an external magnetic field, we add the spin-Zeeman term
\begin{equation}
  \label{eq:Zeeman}
  \mathcal H_{\rm Z}
  =
  -
  \frac{1}{2}
  \sum_i
  \sum_{\mu}
  \sum_{ss'}
  c^{\dagger}_{i\mu s}
  \left(
  \bm H\cdot\bm\sigma
  \right)_{ss'}
  c_{i\mu s'} ,
\end{equation}
where $\bm H$ denotes the effective magnetic field including the factor $-g\mu_{\rm B}$, 
and $\bm\sigma$ is the vector of Pauli matrices.
For each value of $\bm H$, we solve the Hartree-Fock equations
self-consistently and calculate the electric polarization using the
Berry-phase formulation based on the Wilson loop of the occupied Hartree-Fock
eigenstates~\cite{King-Smith1993,Resta1994,Marzari1997}.
Details of the Wilson-loop calculation are provided in
Appendix~\ref{app:polarization_calculation}.
The magnetoelectric response is characterized by the linear relation
\begin{equation}
  \label{eq:magnetoelectric}
  P_\xi
  =
  \alpha_{\xi\xi'} H_{\xi'},
\end{equation}
where $\bm H$ is the effective magnetic field defined above, 
$\bm P$ is the electric polarization, and $\xi,\xi'$ denote the spatial
directions considered here, $\xi,\xi'=a,b,x,y$.
Thus, the slope of the $\bm P$-$\bm H$ curve gives the magnetoelectric
coefficient $\alpha_{\xi\xi'}$.

\begin{figure}[tbp]
  \centering
      \includegraphics[width=0.74\columnwidth,clip]{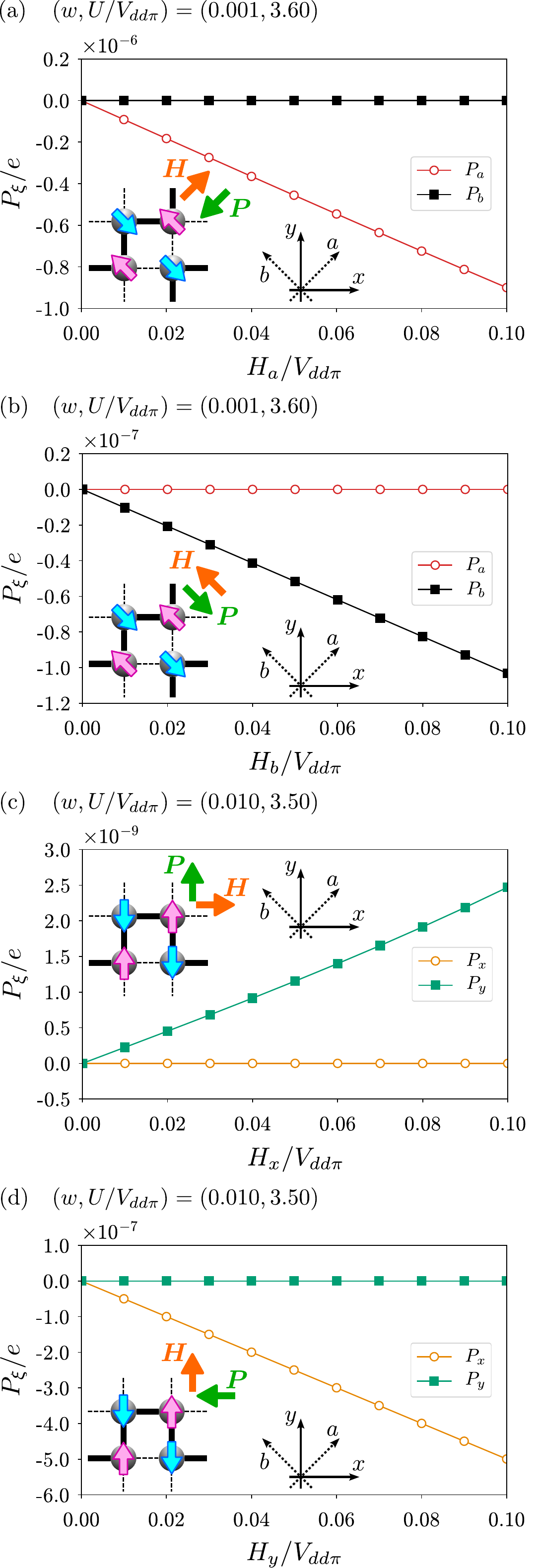}
      \caption{
      Electric polarization induced by the applied magnetic field in the
      AFM+zigzag and AFM+meander phases.
      The spin-orbit coupling is set to $\lambda/V_{dd\pi}=0.05$.
      (a) and (b) show the results for the AFM+zigzag phase at
      $(w,U/V_{dd\pi})=(0.001,3.60)$, where the spin direction is along the $b$ axis.
      (c) and (d) show the results for the AFM+meander phase at
      $(w,U/V_{dd\pi})=(0.010,3.50)$, where the spin direction is along the $y$ axis.
      The insets show the corresponding zero-field spin and bond configurations
      obtained in the presence of spin-orbit coupling.
      The orange and green arrows indicate the directions 
      of the applied magnetic field $\bm H$ and induced polarization $\bm P$, respectively.
      The linear slopes correspond to the magnetoelectric coefficients $\alpha_{\xi\xi'}$.}
      \label{fig:polarization}
\end{figure}

To determine the magnetic easy axis in each phase, the spin-orbit coupling is
set to $\lambda/V_{dd\pi}=0.05$.
The magnetoelectric responses are calculated at
$(w,U/V_{dd\pi})=(0.001,3.60)$ for the AFM+zigzag phase and
$(0.010,3.50)$ for the AFM+meander phase.
For these parameters, the spin direction is selected along the $b$ axis in
the AFM+zigzag phase and along the $y$ axis in the AFM+meander phase.
The corresponding magnetic point groups are $m'm'm'$ and $mmm'$, respectively.

Figure~\ref{fig:polarization} shows the electric polarization induced by the
applied magnetic field in the AFM+zigzag and AFM+meander phases.
The insets show the corresponding zero-field spin and bond configurations.
In the AFM+zigzag phase, $P_a$ and $P_b$ vary linearly with $H_a$ and $H_b$,
respectively, indicating finite diagonal magnetoelectric components.
The response $\alpha_{aa}$ to a magnetic field perpendicular to the easy axis
is approximately one order of magnitude larger than the parallel-field response
$\alpha_{bb}$.
In the AFM+meander phase, $P_y$ and $P_x$ vary linearly with $H_x$ and $H_y$,
respectively, indicating finite off-diagonal magnetoelectric components.
In contrast to the AFM+zigzag phase, the parallel-field response
$\alpha_{xy}$ is much larger than the perpendicular-field response
$\alpha_{yx}$.
These allowed components are consistent with the magnetic point groups of the
two phases.

\subsection{Altermagnetic properties of AFM+tetramer phase}
\label{sec:Altermagnet}

To establish altermagnetism in the AFM+tetramer phase, we examine its
electronic band structure.
We use the representative parameter set
$(w,U/V_{dd\pi})=(0.100,3.60)$ and compare the band structures for
$\lambda/V_{dd\pi}=0.00$ and $0.05$. For $\lambda/V_{dd\pi}=0.05$,
the magnetic easy axis is along the $b$ axis 
as shown in Fig.~\ref{fig:coexistence_phases}(d), 
while for $\lambda=0$,
where the spin direction is degenerate, we choose the same direction
for direct comparison.

\begin{figure}[tbp]
  \centering
      \includegraphics[width=1.0\columnwidth,clip]{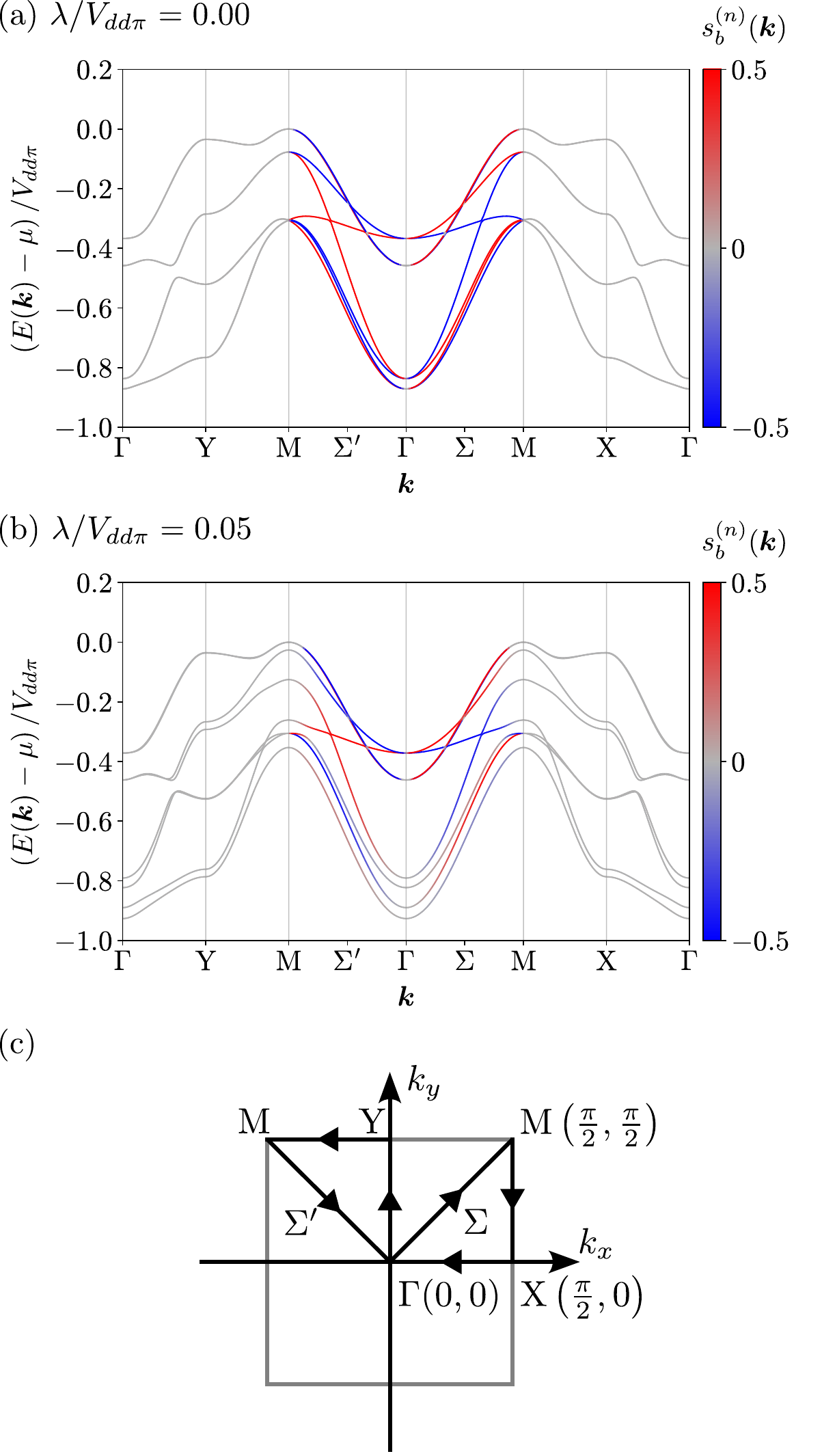}
      \caption{
        Electronic band structures of the valence states mainly derived from the
        $\pi_g$ orbitals in the AFM+tetramer phase
        (a) without spin-orbit coupling, $\lambda/V_{dd\pi}=0.00$, and
        (b) with spin-orbit coupling, $\lambda/V_{dd\pi}=0.05$.
        Both calculations are performed at $(w,U/V_{dd\pi})=(0.100,3.60)$.
        The energies are measured relative to the chemical potential $\mu$, which is
        located at the valence-band maximum.
        The color scale represents the spin expectation value along the $b$ axis,
        $s_b^{(n)}(\bm k)$.
        (c) shows the high-symmetry path in the two-dimensional Brillouin zone
        used in (a) and (b).
        }
      \label{fig:eband}
\end{figure}

Figure~\ref{fig:eband} shows the band structures calculated (a) without and (b) with spin-orbit
coupling.  Because our discussion focuses on the occupied electronic states
near the chemical potential, only the valence bands mainly derived from the
$\pi_g$ orbitals are displayed.  The lower-energy occupied bands mainly derived from the $\sigma_g^z$ orbital and the higher-energy conduction bands are omitted.
For each Bloch eigenstate $|\psi_n(\bm k)\rangle$, we evaluate the spin
expectation value along the $b$ axis as
\begin{equation}
  \label{eq:band_spin_expectation}
  s_b^{(n)}(\bm k)
  =
  \frac{1}{2}
  \langle \psi_n(\bm k) | \sigma^b | \psi_n(\bm k) \rangle.
\end{equation}
The valence bands exhibit a clear spin splitting along
the $\Gamma$-$\mathrm{M}$ lines, and the split branches have opposite signs of
$s_b^{(n)}(\bm k)$, as indicated by the red and blue colors.
Importantly, the spin splitting is already present for $\lambda/V_{dd\pi}=0.00$, as shown in Fig.~\ref{fig:eband}(a), demonstrating its nonrelativistic origin.
It persists in the presence of spin-orbit coupling at $\lambda/V_{dd\pi}=0.05$, as shown in Fig.~\ref{fig:eband}(b).
Thus, the splitting originates from the $\mathcal{PT}$-broken AFM+tetramer order rather than from spin-orbit coupling.

Having established the altermagnetic band character of the AFM+tetramer phase,
we next examine the piezomagnetic response~\cite{Dzialoshinskii1958,Moriya1959,BorovikRomanov1960} as a macroscopic response associated
with this $\mathcal{PT}$-broken AFM state. 
The piezomagnetic effect describes the induction of a magnetization by uniaxial lattice strain. 
In the present calculation, we evaluate the response at finite temperature
using the same parameter set as in the band-structure calculation with
spin-orbit coupling, namely $\lambda/V_{dd\pi}=0.05$ and
$(w,U/V_{dd\pi})=(0.100,3.60)$. 
For these parameters, the magnetic easy axis remains along the $b$ axis even at finite temperatures.

We consider uniaxial pressure along the $a$ or $b$ direction, which
makes the two sets of AFM sublattices carrying opposite spin moments
along the $b$ axis inequivalent.
We incorporate this effect by introducing
an anisotropic modulation of the next-nearest-neighbor hopping matrices,
\begin{equation}
  \label{eq:uniaxial_pressure}
  t^a
  \rightarrow
  (1+p)t^a,
  \qquad
  t^b
  \rightarrow
  (1-p)t^b,
\end{equation}
where $p$ is a dimensionless parameter characterizing the pressure-induced
anisotropy.  The matrices $t^a$ and $t^b$, defined in
Eq.~\eqref{eq:hopping_matrices}, describe the next-nearest-neighbor hoppings
along the $a$ and $b$ directions, respectively.  Their correspondence to the
individual bonds is summarized in Table~\ref{tab:bond_definition}.
A positive value of $p$ enhances the hopping along the $a$ direction and
suppresses that along the $b$ direction, while a negative value reverses this
anisotropy.
Accordingly, $p>0$ and $p<0$ mimic uniaxial compression along the
$a$ and $b$ directions, respectively.

\begin{figure}[tbp]
  \centering
      \includegraphics[width=0.85\columnwidth,clip]{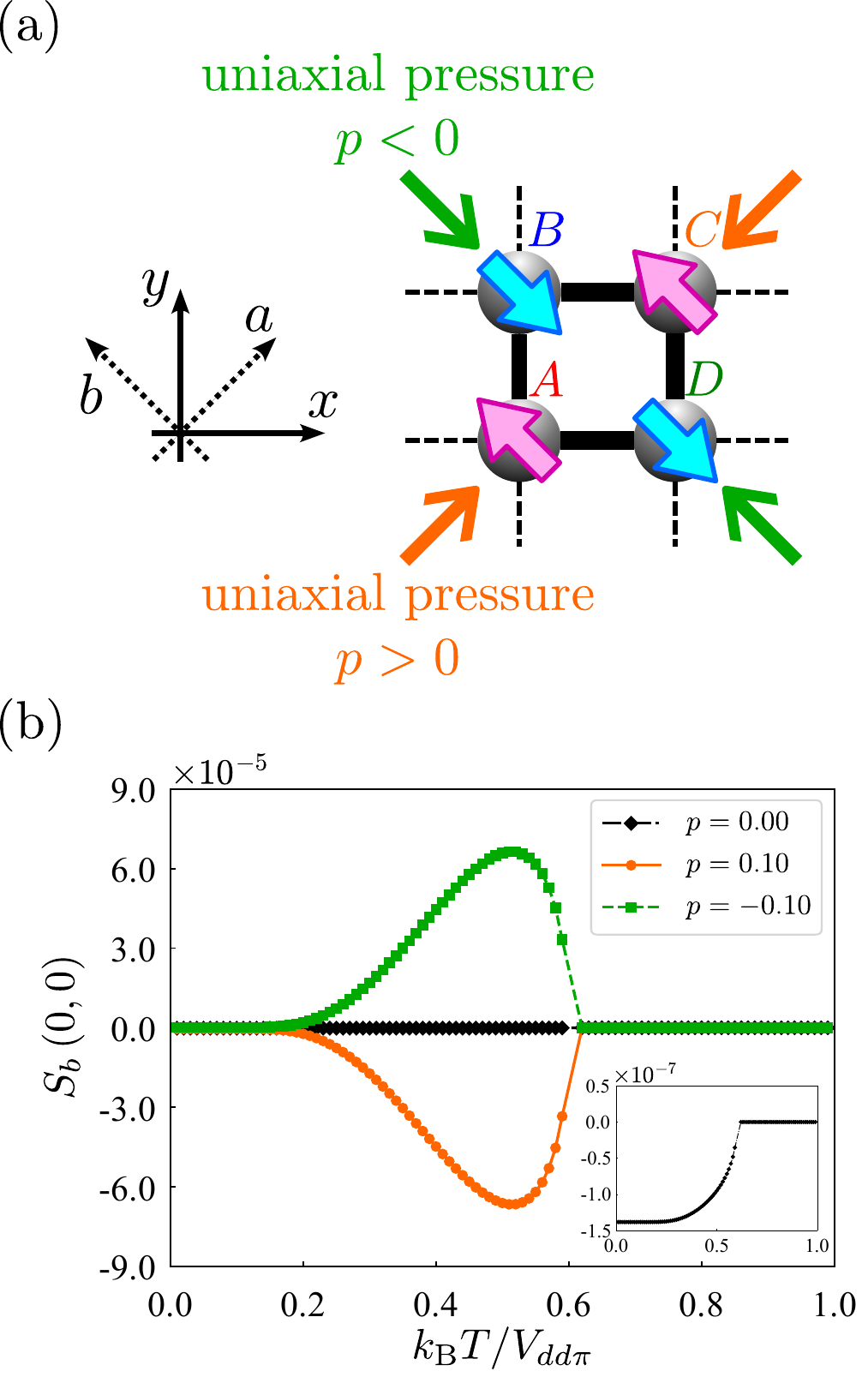}
      \caption{
        Piezomagnetic response of the AFM+tetramer phase 
        under anisotropic next-nearest-neighbor hopping modulation.
        (a) Ground-state spin and bond configuration of the AFM+tetramer phase.
        The orange and green arrows indicate uniaxial pressure along the $a$ and 
        $b$ directions, represented by $p>0$ and $p<0$, respectively.
        (b) Temperature dependence of the ferromagnetic spin component $S_b(0,0)$ 
        for $p=0.00$, $0.10$, and $-0.10$.
        The inset shows an enlarged view of the tiny ferromagnetic component for $p=0.00$.
        The calculations are performed at $\lambda/V_{dd\pi}=0.05$ 
        and $(w,U/V_{dd\pi})=(0.100,3.60)$.
      }
      \label{fig:piezomagnetic}
\end{figure}

Figure~\ref{fig:piezomagnetic} shows the calculated piezomagnetic response of 
the AFM+tetramer phase.
In Fig.~\ref{fig:piezomagnetic}(a), the arrows indicate uniaxial pressure 
along the $a$ axis for $p>0$ and along the $b$ axis for $p<0$.
The spin and bond configuration shown in the same figure represents the ground 
state in the absence of uniaxial pressure.
Figure~\ref{fig:piezomagnetic}(b) shows the temperature dependence of the 
ferromagnetic spin component $S_b(0,0)$ for $p=0.00$, $p=0.10$, and $p=-0.10$.
For $p=0.00$, a tiny ferromagnetic component induced by spin-orbit coupling 
appears along the $b$ axis, as shown in the inset.
When the hopping anisotropy is introduced, a much larger ferromagnetic 
component appears at finite temperature with negative and positive signs for 
$p=0.10$ and $p=-0.10$, respectively.

These finite-temperature ferromagnetic components reflect the thermally activated contribution 
of the spin-split valence bands to the net magnetization, consistent with the nonrelativistic 
mechanism discussed in Refs.~\cite{Yershov2024,Naka2025Piezomagnetism}.
The reversal of the induced ferromagnetic component can be understood from
the imbalance of the sublattice spin moments.
For $p=0.10$, the enhanced hopping between the $A$ and $C$ sublattices
strengthens the AFM interaction between them.
Because the spin moments on the $A$ and $C$ sublattices are parallel in the
magnetic domain shown in Fig.~\ref{fig:piezomagnetic}(a), the enhanced AFM
interaction reduces the magnitudes of their spin moments.
The spin moments on the $B$ and $D$ sublattices therefore become relatively
larger, producing a net magnetization along the negative $b$ direction.
For $p=-0.10$, the enhanced hopping instead strengthens the AFM interaction
between the $B$ and $D$ sublattices.
Because their spin moments are also parallel, the enhanced AFM interaction
reduces the magnitudes of their spin moments.
The spin moments on the $A$ and $C$ sublattices therefore become relatively
larger, producing a net magnetization along the positive $b$ direction.
Thus, switching the direction of uniaxial pressure between the $a$ and $b$
axes reverses the induced magnetization in the AFM+tetramer phase.

\section{Discussion}
\label{sec:Discussion}

Within the present model, the AFM phase can be associated with the
low-pressure $\alpha$ phase, whereas the nonmagnetic tetramer phase corresponds
to the high-pressure $\varepsilon$ phase.
The effect of increasing pressure can be qualitatively represented in the
phase diagram by decreasing $U/V_{dd\pi}$.  Under compression, the
intermolecular hopping $V_{dd\pi}$ is expected to increase more strongly than
the onsite interaction $U$, thereby reducing the effective correlation
strength. 
Pressure may also increase the relative next-nearest-neighbor
hopping strength $w$, because longer-range intermolecular hopping is expected
to be more sensitive to compression.  Therefore, increasing pressure can be
viewed as moving the system approximately from the upper-left region toward
the lower-right region of the phase diagram in Fig.~\ref{fig:coexistence_phases}(a). 
Along this path, the system evolves from
the AFM phase toward the nonmagnetic tetramer phase through an intermediate
region in which AFM and bond orders coexist.
The presence of this intermediate coexistence region reflects the strong
coupling between the magnetic and structural degrees of freedom near the
$\alpha$-$\varepsilon$ transition.

Experimentally, the $\delta$ phase appears between the $\alpha$ and
$\varepsilon$ phases~\cite{Freiman2004,Goncharenko2004}. 
This correspondence suggests that the intermediate coexistence region obtained
in the present model may be associated with the $\delta$ phase. 
Although the present model does not explicitly reproduce the crystal structure of the $\delta$ phase, it suggests that AFM and bond orders may coexist in this intermediate-pressure phase. 
In particular, the AFM+zigzag and AFM+meander states provide possible 
inversion-symmetry-broken magnetic states that can exhibit linear magnetoelectric responses.

Recent single-crystal X-ray diffraction measurements reported an isostructural
transition within the $\varepsilon$ phase at 18.1 GPa and interpreted it as a
structural signature of the spin collapse from the spinful $\varepsilon_1$
state to the spinless $\varepsilon_0$ state~\cite{Gorelli2025}.  In that
interpretation, the $\varepsilon_1$ phase retains molecular spin degrees of
freedom with short-range AFM correlations inside the O$_8$ tetramers, whereas
the $\varepsilon_0$ phase corresponds to a spin-collapsed state.
From this viewpoint, the AFM+tetramer phase obtained in the present model may
be associated with a tetramerized state that retains molecular magnetism. 
Although the Hartree-Fock state does not directly
describe the collective singlet or short-range spin correlations proposed for
the $\varepsilon_1$ phase, it captures the coexistence of tetramerization and
AFM order at the mean-field level. 
Because this phase preserves spatial inversion symmetry but breaks
$\mathcal{PT}$ symmetry, it exhibits altermagnetic spin splitting and a
piezomagnetic response.  Our results therefore suggest that the spinful
tetramerized $\varepsilon_1$ regime could host altermagnetic behavior in
high-pressure solid oxygen before the complete spin collapse into the spinless
$\varepsilon_0$ state.

\section{Summary}
\label{sec:Summary}

We investigated the interplay between magnetic and structural orders in solid
oxygen using a three-orbital Peierls--Hubbard model treated within the
Hartree-Fock approximation.  The calculated phase diagram contains an AFM
phase, three bond-ordered phases with zigzag, meander, and tetramer
distortions, and three coexistence phases of AFM and bond orders.  With
decreasing effective correlation strength, the AFM phase evolves into a
bond-ordered phase through an intermediate coexistence region, while the type
of bond order is controlled by the relative next-nearest-neighbor hopping
strength.
The symmetry and electronic responses of the coexistence phases depend on the
bond-distortion pattern.  The AFM+zigzag and AFM+meander phases break spatial
inversion symmetry and exhibit linear magnetoelectric responses.  In contrast,
the AFM+tetramer phase breaks $\mathcal{PT}$ symmetry and exhibits
altermagnetic spin splitting even without spin-orbit coupling.  An anisotropic
modulation of the next-nearest-neighbor hopping also induces a piezomagnetic
response in this phase.  These findings demonstrate that the coexistence of
AFM order and bond distortion can stabilize magnetic phases
whose broken symmetries give rise to distinct electronic responses.

Although the present analysis is restricted to two dimensions, oxygen molecules form three-dimensional crystal structures in solid oxygen.
This implies additional degrees of freedom associated with the relative arrangement of spin and bond configurations in adjacent layers, which may affect the crystal and magnetic symmetries.
Indeed, neutron diffraction measurements have revealed temperature-dependent changes in the AFM stacking sequence within the $\delta$ phase~\cite{Klotz2010}.
This finding motivates further investigation of whether different stacking patterns of the coexistence phases identified on a square lattice can yield nonzero macroscopic magnetoelectric or piezomagnetic responses in three dimensions.
Moreover, the Hartree-Fock approximation used here describes coexistence phases with long-range AFM order but neglects quantum fluctuations.
Since the molecular magnetic moments have been proposed to form a collective singlet state with short-range AFM correlations in the $\varepsilon_1$ phase~\cite{Crespo2014}, an approach capable of including quantum fluctuations is needed to go beyond the present mean-field description.
Previous quantum Monte Carlo studies have found tetramer-type bond order in a related Peierls--Heisenberg model on a square lattice~\cite{Aits2006}.
Applying such a method to an effective model derived by considering the strong correlation limit could be a possible direction for future work.

\begin{acknowledgments}
The authors thank T. Aoyama and T. Kimura for fruitful discussions.
This work was supported by Grants-in-Aid for Scientific Research from JSPS,
KAKENHI Grant Nos. JP20H00122, JP22H01175, JP23K25826, JP23H04865, JP24K00563, 
and JP26H00624.
\end{acknowledgments}

\appendix
\section{Orbital angular momentum in the molecular-orbital basis}
\label{app:orbital_angular_momentum}

Here, we derive the orbital angular momentum matrices given in
Eq.~\eqref{eq:orbital_angular_momentum} for the three molecular orbitals
retained in the present model, $|\pi_g^x\rangle$, $|\pi_g^y\rangle$, and
$|\sigma_g^z\rangle$.
For compact notation, we denote these orbitals by $|x_g\rangle$,
$|y_g\rangle$, and $|z_g\rangle$, respectively.
The two oxygen atoms in an ${\rm O_2}$ molecule are labeled $A$ and $B$.
The corresponding atomic $2p$ orbital wave functions are represented by
$|p_\mu,A\rangle$ and $|p_\mu,B\rangle$, where the orbital
index $\mu$ runs over $x$, $y$, and $z$.
Within the linear combination of atomic orbitals (LCAO) approximation, these
molecular orbitals are formed as linear combinations of the atomic $2p$
orbitals on atoms $A$ and $B$. 
Neglecting the interatomic overlap in the normalization, 
we write the molecular orbital wave functions as
\begin{equation}
  |\mu_g\rangle
  =
  \frac{1}{\sqrt{2}}
  \left(
  |p_\mu,A\rangle-|p_\mu,B\rangle
  \right).
\end{equation}

We employ the one-center approximation for the orbital angular momentum
operator.  In this approximation, we neglect the interatomic matrix elements,
\begin{equation}
  \langle p_\mu,A|\ell^\alpha|p_\nu,B\rangle
  =
  \langle p_\mu,B|\ell^\alpha|p_\nu,A\rangle
  \simeq 0.
\end{equation}
Accordingly, the matrix elements in the molecular orbital basis reduce to
\begin{align}
  \langle \mu_g|\ell^\alpha|\nu_g\rangle
  &\simeq
  \frac{1}{2}
  \sum_{\tau=A,B}
  \langle p_\mu,\tau|\ell^\alpha|p_\nu,\tau\rangle \\
  &=
  \langle p_\mu|\ell^\alpha|p_\nu\rangle,
  \label{eq:mo_angular_momentum_projection}
\end{align}
where we have used the fact that the onsite matrix elements on atoms $A$ and
$B$ are identical.
In units of $\hbar$, the angular momentum matrix elements in the atomic $p$
orbital basis are given by
\begin{equation}
  \langle p_\mu|\ell^\alpha|p_\nu\rangle
  =
  -i\epsilon_{\alpha\mu\nu},
\end{equation}
where $\epsilon_{\alpha\mu\nu}$ is the Levi-Civita symbol.
Consequently, the representation of $\bm\ell$ in the molecular orbital basis
$\{|\pi_g^x\rangle,|\pi_g^y\rangle,|\sigma_g^z\rangle\}$ is identical to its
standard representation in the atomic basis
$\{|p_x\rangle,|p_y\rangle,|p_z\rangle\}$, yielding the three matrices in
Eq.~\eqref{eq:orbital_angular_momentum}.

The neglected interatomic spin-orbit contributions are expected to be small
owing to cancellation between terms with opposite signs.
As a quantitative benchmark, the same one-center approximation yields a
spin-orbit coupling constant of $-153\,\mathrm{cm}^{-1}$ for ${\rm O_2^-}$,
which is close to the experimental value of $-160\,\mathrm{cm}^{-1}$, with a
difference of about $4\%$~\cite{Prabhakar2002}.

\section{Wilson-loop calculation of electric polarization}
\label{app:polarization_calculation}

Here, we provide details of the Wilson-loop calculation used to evaluate the
electric polarization in Sec.~\ref{sec:MagnetoelectricResponse}.
We write the polarization in the two-dimensional plane as
$\bm P=P_x\hat{\bm e}_{x}+P_y\hat{\bm e}_{y}$ and describe below the
calculation of $P_x$.
The Brillouin zone is discretized into an $N_x\times N_y$ mesh,
\begin{equation}
  \label{eq:k_mesh}
  k_x^\ell
  =
  \pi \left(\frac{\ell}{N_x}-\frac{1}{2}\right),
  \qquad
  k_y^j
  =
  \pi \left(\frac{j}{N_y}-\frac{1}{2}\right),
\end{equation}
where $\ell=0,\ldots,N_x-1$ and $j=0,\ldots,N_y-1$.
For each fixed $k_y^j$, we construct the Wilson loop along $k_x$ as
\begin{equation}
  \label{eq:wilson_loop_x}
  W_x(k_y^j)
  =
  \prod_{\ell=0}^{N_x-1}
  M_x(k_x^\ell,k_x^{\ell+1};k_y^j),
\end{equation}
where $k_x^{N_x}\equiv k_x^0$ due to the periodicity of the Brillouin zone.
The overlap matrix is defined as
\begin{equation}
  \label{eq:wilson_matrix_x}
  \left[
  M_x(k_x^\ell,k_x^{\ell+1};k_y^j)
  \right]_{mn}
  =
  \langle
  \psi_{m}(k_x^\ell,k_y^j)
  |
  \psi_{n}(k_x^{\ell+1},k_y^j)
  \rangle,
\end{equation}
where $\psi_{n}(k_x^\ell,k_y^j)$ is an occupied single-particle eigenstate of
the self-consistent Hartree-Fock Hamiltonian, and $m$ and $n$ label the
occupied bands.
When the link crosses the Brillouin-zone boundary, namely for
$\ell=N_x-1$, the overlap matrix is evaluated by relating
$\psi_n(k_x^{N_x},k_y^j)$ to $\psi_n(k_x^0,k_y^j)$ with the boundary phase
factor associated with the basis positions in the unit cell~\cite{Vanderbilt2018}.
The Berry phase is obtained as
\begin{equation}
  \label{eq:berry_phase_x}
  \phi_x(k_y^j)
  =
  -{\rm Im}\ln\det W_x(k_y^j).
\end{equation}
The polarization along the $x$ direction is then evaluated by averaging the
Berry phase over $k_y$,
\begin{equation}
  \label{eq:polarization_x}
  P_x
  =
  -
  \frac{e}{\pi N_y}
  \sum_{j=0}^{N_y-1}
  \phi_x(k_y^j),
\end{equation}
where $e>0$ is the elementary charge and the nearest-neighbor intermolecular
distance is set to unity.
The polarization is expressed in this model unit.
The factor $\pi$ in the denominator reflects the choice of the four-sublattice
unit cell, whose period is twice the nearest-neighbor distance along the
corresponding direction.
The polarization $P_y$ is calculated in a similar manner to $P_x$.
The components along the crystallographic $a$ and $b$ axes are obtained from
$P_x$ and $P_y$ by the corresponding coordinate transformation.

\FloatBarrier

\bibliography{refs}
\end{document}